\documentclass[english]{article}
\usepackage[T1]{fontenc}
\usepackage[latin9]{inputenc}
\usepackage{geometry}
\usepackage{amsmath}
\usepackage{amssymb}
\usepackage{graphicx}

\makeatletter

\providecommand{\tabularnewline}{\\}

\@ifundefined{date}{}{\date{}}
\makeatother

\usepackage{babel}
\begin{document}
\title{\textbf{A Descriptive Method of Firm Size Transition Dynamics Using
Markov Chain}}
\maketitle
\begin{center}
{\large{}Boyang You}\footnote{Ph.D. Student, Department of Economics, University of Bath. Email:
B.You@bath.ac.uk}{\large{}, Kerry Papps}\footnote{Senior Lecturer, Department of Economics, University of Bath. Centre
for Analysis of Social Policy (CASP) Labour, Education and Health
Economics. Email: K.L.Papps@bath.ac.uk}{\large\par}
\par\end{center}

$ $

$ $
\begin{abstract}
Social employment, which is mostly carried by firms of different types,
determines the prosperity and stability of a country. As time passing,
the fluctuations of firm employment can reflect the process of creating
or destroying jobs. Therefore, it is instructive to investigate the
firm employment (size) dynamics. Drawing on the firm-level panel data
extracted from the Chinese Industrial Enterprises Database 1998-2013,
this paper proposes a Markov-chain-based descriptive approach to clearly
demonstrate the firm size transfer dynamics between different size
categories. With this method, any firm size transition path in a short
time period can be intuitively demonstrated. Furthermore, by utilizing
the properties of Markov transfer matrices, the definition of transition
trend and the transition entropy are introduced and estimated. As
a result, the tendency of firm size transfer between small, medium
and large can be exactly revealed, and the uncertainty of size change
can be quantified. Generally from the evidence of this paper, it can
be inferred that small and medium manufacturing firms in China have
greater job creation potentials compared to large firms over this
time period.

$ $
\end{abstract}


$ $

\section{Introduction}

Firms are business organizations which carrying the vast majority
of social employment. When defining the size of firms as their number
of employees (Birch, 1979), the net job creation can be calculated
by the firm size variations. Early studies mostly focus on quantifying
net job creation of firms via cross-sectional data sets: Birch (1979,
1981, 1987), Davis \& Haltiwanger (1994), Neumark et al. (2011) all
look into the job creation abilities of different firm types including
small, medium and large firms.

In practice, the process of firm job creation is continuous. The changes
of firm job creation in various time periods can be regarded as the
dynamics of firm size, making scholars examine this issue via time
series approaches. One of the classic method used is the panel vector
autoregression (PVAR, Love \& Zicchino, 2006). By this model, the
impact among firm characteristics can be clearly estimated. Davis
\& Haltiwanger (1999, 2001), Koellinger \& Roy (2012) all deeply examine
the firm employment or entrepreneurship determinants via this method.

The PVAR method holds the advantages of expressing how the dynamic
changes of individual firms are influenced by its own. However, this
approach cannot clearly quantify the net job creation of firms in
different size categories as Birch's studies do. Moreover, data stationary
and data time length is strictly required when using autoregressive
models. Due to the existence of firm entry \& exit, existing firm-level
data sets can hardly meet the requirement. This inspires us to further
propose a Markov-chain-based approach to describe the dynamics of
firm size.

Markov chain, which can describe the random transfer processes is
widely used in many aspects. Unlike autoregression models, any two-period
transfer can be described by establishing transition matrix. Using
the properties of Markov chain, multiple first-order transfer matrices
can also describe the dynamics over a continuous time period in an
intuitive way. In the topic of firms, there are currently not many
relevant studies. Uyar (1972) forecasts the replacement demand of
employees using Markov chain. Horowitz \& Horowitz (1968) use a first-order
Markov process to describe market change. Joining Markov first-order
transfer matrices with entropy, the industrial agglomeration of the
brewing industry is quantified. Kopecky \& Suen (2010) propose a method
that can use Markov chain to approximate autoregression models.

In this paper, a method that describes firm size transition dynamics
is proposed based on Markov chain. Joint with a rich continuous panel
data exacted from the Chinese Industrial Enterprises Database 1998-2013,
the first order transfer matrices are firstly calculated, and the
dynamic size transfer path of Chinese manufacturing firms between
1998 to 2013 are revealed. Furthermore, using the properties of Markov
Chain, not only the job creation amount but also the general firm
growth trend of each can be presented by calculating the probability
of upward or downward transfer in the matrices. Besides, following
the idea of Horowitz \& Horowitz (1968), entropy of firm size are
defined and measured. Through these results, the job creation probability
of firms in any time period or any size category can be directly described,
which may intuitively help policy maker recognize firms with which
size categories may create more jobs, and in turn formulate effective
employment-boosting polices for the target firms to alleviate the
employment pressure of the society in an effective way.

For the rest of this paper, section 2 illustrates the methodology
of the proposed method, in which the definition of firm size transition
and the Markov first-order transition matrix is given. Based on the
transition matrices, the transfer path, trend and entropy are defined.
Section 3 shows the results of using the proposed method over a firm-level
data set, which includes data set descriptions, calculations details,
results analyses and robustness check. Section 4 is the conclusion
and discussion of this paper.

\section{Methodology}

In this section, the methodology of the transition analysis will be
presented. First, the size of the firm is divided into multiple states,
which is judged by the number of employees of firms from small to
large. Next, existing panel data will be rectangularized including
firm entry and exit, and the Markov state transfer matrices between
different consecutive years will be calculated. By analysing the obtained
Markov matrices, a general trend of inter-transition in terms of firm
size may be concluded, while the average job creation among all categories
will be calculated based on transfer matrices.

\subsection{Definition of firm size transition}

The basic idea of using Markov chain is to illustrate how the size
of firms, which belongs to one of the $N$ categories, are relocated
after $d$ years. The entire process of the transfer constitutes multiple
probability transition matrices which can be considered as a Markov
chain, while all these matrices can be calculated from the sample
data set. 

Based on this idea, it is necessary to find that among all firms,
how many of them has transferred from size category $i$ to $j$ after
$d$ years. To this end, let $f_{ji}(D,D+d)$ be the possibility of
the size of a firm transfers from category $i$ to $j$, from base
year $D$ to year $D+d$. Let the categories in terms of firm size
at year $d$ be $S_{d},\:S_{d}\in\left\{ 0,1,2,\text{\dots},N\right\} $.
Then there is:
\begin{equation}
f_{ji}(D,D+d)=Pr\left(S_{D+d}=j\mid S_{D}=i\right)\;\;\;\;\;i,j\in\left\{ 0,1,2,\text{\dots},N\right\} 
\end{equation}

From equation (1), it is obvious to find that the firm size transition
process can be described as a Markov chain. Also, for notation simplicity,
in this paper, assuming base year $D=0$, which means the base year
the firm size transfer begins at year $0$. Hence, there is $f_{ji}(D,D+d)=f_{ji}(0,d)$.

\subsection{Properties of Markov chain}

According to Neumark, et al. (2011), firms are divided into 13 size
categories according to their employee numbers, in which category
0 represents the non-existing firms at current year. Similar to You
\& Fan (2020), let $N=12$. Then, the transfer matrix of the firm
size can be considered as a $[13\times13]$ matrix. Let $F(d-1,d)$
be a first order 13-state Markov transfer matrix. Based on equation
(2), there is:

\begin{equation}
F(d-1,d)=\left[\begin{array}{cccc}
f_{00}(d-1,d) & f_{01}(d-1,d) & \cdots & f_{0N}(d-1,d)\\
f_{10}(d-1,d) & f_{11}(d-1,d) & \cdots & f_{1N}(d-1,d)\\
\vdots & \vdots &  & \vdots\\
f_{N0}(d-1,d) & f_{N1}(d-1,d) & \cdots & f_{NN}(d-1,d)
\end{array}\right],
\end{equation}

where elements in every column satisfies:
\begin{equation}
\sum_{j=0}^{N}f_{ji}(d-1,d)=1
\end{equation}

Note that $F(d-1,d)$ tend to be a diagonally dominant matrix if firm
entry and exit (state 0) are excluded. This is because, for most companies,
size change between two consecutive years is relatively small. That
is to say, for any $i,j=0,1,2,\ldots,N,j\neq i$, roughly satisfying
$f_{ii}(d-1,d)\geqslant f_{ji}(d-1,d)$.

Also, according to the Law of total probability, there is:

\begin{equation}
Pr\left(S_{d}=j\right)=\sum_{i=0}^{N}Pr\left(S_{d}=j\mid S_{d-1}=i\right)Pr\left(S_{d-1}=i\right)
\end{equation}

Let $p_{j}(d)=Pr\left(S_{d}=j\right)$, then, there is:

\begin{equation}
p_{j}(d)=\sum_{i=0}^{N}f_{ji}(d-1,d)p_{i}(d-1)
\end{equation}

\subsubsection*{Transition path}

The Gibrat Law indicates that the evolution of the firm size is only
related to the size of its previous year. Following this idea, a transition
path can be described by a Markov chain. Let $P(d)=\left[\begin{array}{cccc}
p_{0}(d) & p_{1}(d) & \cdots & p_{N}(d)\end{array}\right]$$^{'}$. For any $d$ there is:

\begin{equation}
P(d)=F(d-1,d)P(d-1)
\end{equation}

Equation (6) can also be transformed as follows:
\begin{align}
P(d) & =F(d-1,d)P(d-1)\nonumber \\
 & =F(d-1,d)F(d-2,d-1)P(d-2)\nonumber \\
 & =\cdots\nonumber \\
 & =F(d-1,d)F(d-2,d-1)\cdots F(1,2)F(0,1)P(0)
\end{align}

In a special case when the Markov chain is homogeneous, define $F$
which satisfies $F=F(d-1,d)=F(d-2,d-1)=\cdots=F(0,1)$, there is:

\begin{equation}
P(d)=F^{d}P(0)
\end{equation}

It should be noted that the assumption given here is strong. In practice,
the transfer matrices obtained from sample data can hardly be homogeneous.

\subsubsection*{Transition trend}

By analysing the transfer matrices, not only the probability of small
firms evolving into large firms, but also the transition trend can
be discovered. For example, when prosperity comes, firms may be willing
to increase recruitment, which will make the distribution of the transfer
matrix elements move to the bottom left corner; conversely, when recession
comes, firms may reduce recruitment or even layoffs, leading to the
distribution of the transfer matrix elements moving to the upper right
corner. 

In order to quantify this trend, let 
\begin{equation}
R(d)=\sum_{j<i}f_{ji}(d-1,d)p_{j}(d)
\end{equation}
 
\begin{equation}
L(d)=\sum_{j>i}f_{ji}(d-1,d)p_{j}(d)
\end{equation}
Next, simply examine $Q(d)=\frac{L(d)}{R(d)}$, the transition trend
at year $d$ can be described.

\subsubsection*{Transition entropy}

In the study of Horowitz \& Horowitz (1968), the concept of entropy
in information theory, which may reveal the uncertainty, are used
to quantify industrial agglomeration. Following this idea, it is proper
to introduce a concept of transition entropy, which reveals the randomness
of firm size evolution into out topic. With this, the likelihood of
firm size change in a certain category can be quantified and compared.

According to the definition proposed by Shannon (1948), the entropy
of the $i-th$ column element of the transfer matrix, for each $f_{ji}(d-1,d)>0$,
can be expressed as:

\begin{equation}
I_{i}(d)=-\sum_{j=0}^{N}f_{ji}(d-1,d)\cdot\ln f_{ji}(d-1,d),
\end{equation}

where $I_{i}(d)$ is the entropy of the $i-th$ firm category. Note
that the greater the entropy of the $i-th$ firm category, the greater
the randomness of its size change.

\section{Evaluations over the data set}

\subsection{Data}

The data set used in this paper is extracted from the Chinese Industrial
Enterprises Database 1998-2013. Variable SIZE is defined by the average
number of employees of firms in the year. The descriptive statistics
of the data set used in this paper is shown in Table 1.
\noindent \begin{center}
Table 1 Descriptive statistics of Markov Transition section
\par\end{center}

\noindent \begin{center}
\begin{tabular}{lllllll}
\hline 
{\scriptsize{}Variable} & {\scriptsize{}Obs.} & {\scriptsize{}Variable Unit} & {\scriptsize{}Mean} & {\scriptsize{}Min} & {\scriptsize{}Max} & {\scriptsize{}Std. Dev.}\tabularnewline
\hline 
{\scriptsize{}SIZE} & {\scriptsize{}3,747,157} & {\scriptsize{}Person} & {\scriptsize{}293.63} & {\scriptsize{}1} & {\scriptsize{}760,884} & {\scriptsize{}1,453.6}\tabularnewline
\hline 
\multicolumn{7}{r}{{\tiny{}Data source: Chinese Industrial Enterprises Database.}}\tabularnewline
\end{tabular}
\par\end{center}

The data set is selected unbalanced from 1998-2013 due to firm entry
and exit. The sample size in this section is 3,747,157, with its minimum
value 1, maximum value 760,884 and standard deviation 1453.6. For
a purpose of calculation, the unbalanced data set will be further
rectangularized to a balanced data set by filling 0 into missing values,
which are considered as non-existing firms.

\subsection{Calculation details}

First, referencing the firm size classification criterion used in
our previous study (You \& Fan, 2020, Table 1), firms are divided
into size categories of 0-13 with employee number boundaries of 0,
20, 50, 100, 250, 500, 1,000, 2,500, 5,000, 10,000, 25,000 and 50,000.

Second, according to Section 2.2, transfer matrices $F(d-1,d)$ are
built based on the probabilities $f_{ji}(d-1,d),\:i,j\in\left\{ 0,1,2,\text{\dots},12\right\} $.
To calculate each $f_{ji}(d-1,d)$, according to the Borel's law of
large numbers, when sample size is sufficiently large, there is:

\begin{equation}
\frac{Num_{ji}(d)}{Num_{i}(d-1)}\rightarrow^{p}Pr\left(S_{d}=j\mid S_{d-1}=i\right)=f_{ji}(d-1,d),
\end{equation}

where $Num_{i}(d-1)$ is the number of the firms whose size is in
category $i$ at year $d-1$, while $Num_{ji}(d)$ is the number of
firms whose size category changes from category $i$ at year $d-1$
to category $j$ at year $d$.

With this, a Markov chain that reflects dynamic transitions between
firm size categories can be established.

\subsection{Results}

As the sample data is collected between 1998 and 2013, there are 15
first-order transfer matrices obtained, which are defined in equation
(2) while calculated by equation (12). Table 2 is the first-order
transfer matrix from 1998 to 1999.
\noindent \begin{center}
Table 2 The first-order transfer matrix from 1998 to 1999
\par\end{center}

\noindent \begin{center}
{\tiny{}}%
\begin{tabular}{llllllllllllll}
\hline 
{\tiny{}Size} & {\tiny{}0} & {\tiny{}1} & {\tiny{}2} & {\tiny{}3} & {\tiny{}4} & {\tiny{}5} & {\tiny{}6} & {\tiny{}7} & {\tiny{}8} & {\tiny{}9} & {\tiny{}10} & {\tiny{}11} & {\tiny{}12}\tabularnewline
\hline 
{\tiny{}0} & {\tiny{}0.0000} & {\tiny{}0.3688} & {\tiny{}0.2768} & {\tiny{}0.2173} & {\tiny{}0.1743} & {\tiny{}0.1366} & {\tiny{}0.1186} & {\tiny{}0.1219} & {\tiny{}0.1061} & {\tiny{}0.1024} & {\tiny{}0.0815} & {\tiny{}0.0926} & {\tiny{}0.0286}\tabularnewline
{\tiny{}1} & {\tiny{}0.0546} & {\tiny{}0.3947} & {\tiny{}0.0526} & {\tiny{}0.0122} & {\tiny{}0.0055} & {\tiny{}0.0029} & {\tiny{}0.0014} & {\tiny{}0.0017} & {\tiny{}0.0000} & {\tiny{}0.0000} & {\tiny{}0.0000} & {\tiny{}0.0000} & {\tiny{}0.0000}\tabularnewline
{\tiny{}2} & {\tiny{}0.1759} & {\tiny{}0.1099} & {\tiny{}0.4844} & {\tiny{}0.0914} & {\tiny{}0.0163} & {\tiny{}0.0068} & {\tiny{}0.0052} & {\tiny{}0.0018} & {\tiny{}0.0019} & {\tiny{}0.0000} & {\tiny{}0.0000} & {\tiny{}0.0000} & {\tiny{}0.0000}\tabularnewline
{\tiny{}3} & {\tiny{}0.2425} & {\tiny{}0.0455} & {\tiny{}0.1223} & {\tiny{}0.5370} & {\tiny{}0.0921} & {\tiny{}0.0165} & {\tiny{}0.0077} & {\tiny{}0.0024} & {\tiny{}0.0019} & {\tiny{}0.0000} & {\tiny{}0.0043} & {\tiny{}0.0000} & {\tiny{}0.0000}\tabularnewline
{\tiny{}4} & {\tiny{}0.2952} & {\tiny{}0.0566} & {\tiny{}0.0421} & {\tiny{}0.1235} & {\tiny{}0.6403} & {\tiny{}0.1501} & {\tiny{}0.0336} & {\tiny{}0.0091} & {\tiny{}0.0025} & {\tiny{}0.0017} & {\tiny{}0.0000} & {\tiny{}0.0000} & {\tiny{}0.0000}\tabularnewline
{\tiny{}5} & {\tiny{}0.1279} & {\tiny{}0.0160} & {\tiny{}0.0168} & {\tiny{}0.0121} & {\tiny{}0.0628} & {\tiny{}0.6264} & {\tiny{}0.1542} & {\tiny{}0.0254} & {\tiny{}0.0025} & {\tiny{}0.0068} & {\tiny{}0.0000} & {\tiny{}0.0000} & {\tiny{}0.0000}\tabularnewline
{\tiny{}6} & {\tiny{}0.0612} & {\tiny{}0.0029} & {\tiny{}0.0037} & {\tiny{}0.0054} & {\tiny{}0.0066} & {\tiny{}0.0550} & {\tiny{}0.6349} & {\tiny{}0.1404} & {\tiny{}0.0181} & {\tiny{}0.0017} & {\tiny{}0.0000} & {\tiny{}0.0000} & {\tiny{}0.0000}\tabularnewline
{\tiny{}7} & {\tiny{}0.0313} & {\tiny{}0.0056} & {\tiny{}0.0010} & {\tiny{}0.0010} & {\tiny{}0.0018} & {\tiny{}0.0042} & {\tiny{}0.0418} & {\tiny{}0.6703} & {\tiny{}0.1685} & {\tiny{}0.0137} & {\tiny{}0.0000} & {\tiny{}0.0000} & {\tiny{}0.0000}\tabularnewline
{\tiny{}8} & {\tiny{}0.0069} & {\tiny{}0.0000} & {\tiny{}0.0001} & {\tiny{}0.0001} & {\tiny{}0.0003} & {\tiny{}0.0013} & {\tiny{}0.0017} & {\tiny{}0.0247} & {\tiny{}0.6648} & {\tiny{}0.1758} & {\tiny{}0.0043} & {\tiny{}0.0185} & {\tiny{}0.0286}\tabularnewline
{\tiny{}9} & {\tiny{}0.0028} & {\tiny{}0.0000} & {\tiny{}0.0001} & {\tiny{}0.0001} & {\tiny{}0.0000} & {\tiny{}0.0001} & {\tiny{}0.0005} & {\tiny{}0.0018} & {\tiny{}0.0293} & {\tiny{}0.6792} & {\tiny{}0.1631} & {\tiny{}0.0000} & {\tiny{}0.0000}\tabularnewline
{\tiny{}10} & {\tiny{}0.0012} & {\tiny{}0.0000} & {\tiny{}0.0000} & {\tiny{}0.0000} & {\tiny{}0.0000} & {\tiny{}0.0000} & {\tiny{}0.0002} & {\tiny{}0.0006} & {\tiny{}0.0044} & {\tiny{}0.0188} & {\tiny{}0.7253} & {\tiny{}0.2222} & {\tiny{}0.0000}\tabularnewline
{\tiny{}11} & {\tiny{}0.0003} & {\tiny{}0.0000} & {\tiny{}0.0000} & {\tiny{}0.0000} & {\tiny{}0.0000} & {\tiny{}0.0000} & {\tiny{}0.0002} & {\tiny{}0.0000} & {\tiny{}0.0000} & {\tiny{}0.0000} & {\tiny{}0.0215} & {\tiny{}0.6481} & {\tiny{}0.2571}\tabularnewline
{\tiny{}12} & {\tiny{}0.0002} & {\tiny{}0.0000} & {\tiny{}0.0000} & {\tiny{}0.0000} & {\tiny{}0.0000} & {\tiny{}0.0000} & {\tiny{}0.0000} & {\tiny{}0.0000} & {\tiny{}0.0000} & {\tiny{}0.0000} & {\tiny{}0.0000} & {\tiny{}0.0185} & {\tiny{}0.6857}\tabularnewline
\hline 
\multicolumn{14}{r}{{\tiny{}Axis coordinates represent size categories, transfer probabilities
in the matrix. Full matrices see Appendix}}\tabularnewline
\end{tabular}{\tiny\par}
\par\end{center}

As it is imagined before calculating, if firm entry and exit are excluded,
these matrices tend to be diagonally dominant matrices, with most
of non-zero values on the diagonal of the matrices. \footnote{See Appendix for all 15 matrices from 1998 to 2013.}

Next, based on the equations in Section 2.3, transition trend and
transition entropy are calculated.\footnote{See Appendix for full results.}

Figure 1 is the obtained transition trend based on equation (9) and
(10). The value of $Q(d)$ indicated a general transition trend at
year $d$. It can be observed that $Q(d)$ in 2008, 2009 and 2013
are relatively low. This indicates that there is more job destruction
happened in these two years. In contrast, $Q(d)$ values in 2004,
2010 and 2012 are relatively high, which means job creation is far
more prevalent than job destruction in these years. The massive job
destruction in 2008 and 2009 might be attributed to the subprime crisis,
while the massive job creation 2004 and 2012 might be attributed to
national policy changes. Note that the job destruction trend in 2010
shall be attributed to the criteria change of the database, which
no longer include firms with sales less than 20 million CNY instead
of 5 million.
\begin{center}
\includegraphics[width=12cm]{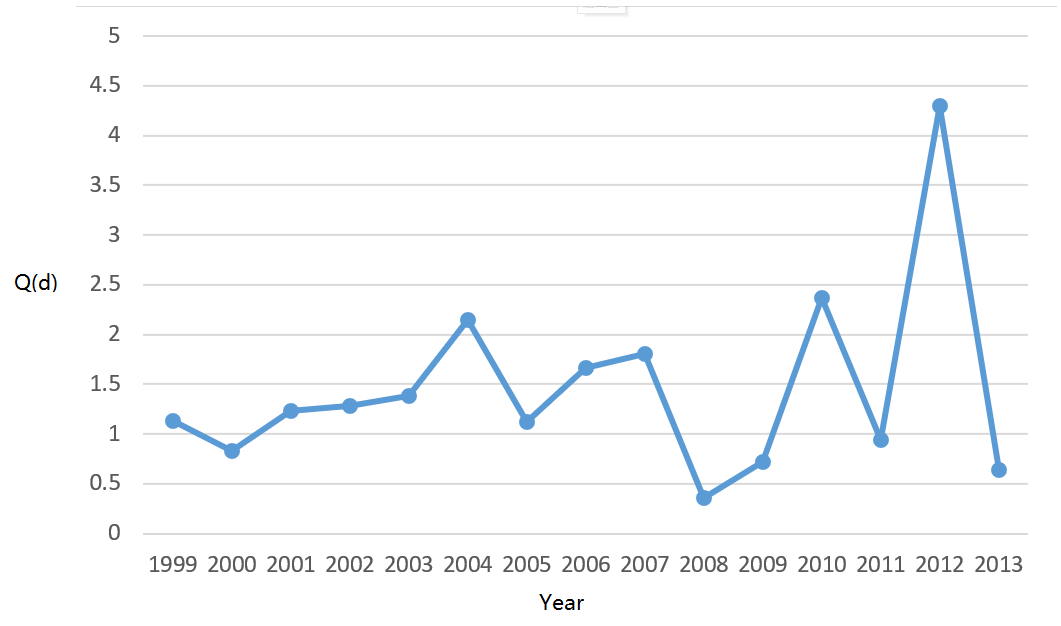}
\par\end{center}

\noindent \begin{center}
Figure 1 The transition trend from 1999 to 2013
\par\end{center}

\noindent \begin{center}
{\tiny{}Year means end year. e.g., year 1999 represents 1998 to 1999.
$Q(d)$ is transition trend. Full data see Appendix.}{\tiny\par}
\par\end{center}

$ $

Figure 2 is the obtained empirical result based on equation (11).
In the figure firms in categories 1-3 are defined as small firms,
firms in categories 4-6 are defined as mediums firms and firms in
categories 7-12 are defined as large firms. From different size categories
it can be observed that small firms tend to have greater transfer
entropy while large firms tend to have smaller transfer entropy. The
entropy of small firms fluctuates around 1.2. For the entropy of medium
firms, there is a rough entropy difference of 0.1 in average compared
to small firms. The entropy of large firms have the lowest value in
average, which fluctuates around 1. From the perspective over time
it can be observed that in 2001, 2004, 2008 and 2010, the transition
entropy values increase significantly, which means the likelihood
of size transfer increase. This might be attributed to the instability
of industrial economic environment. Note that in 2012, due to missing
data, the transition entropy cannot be accurately estimated.
\begin{center}
\includegraphics[width=12cm]{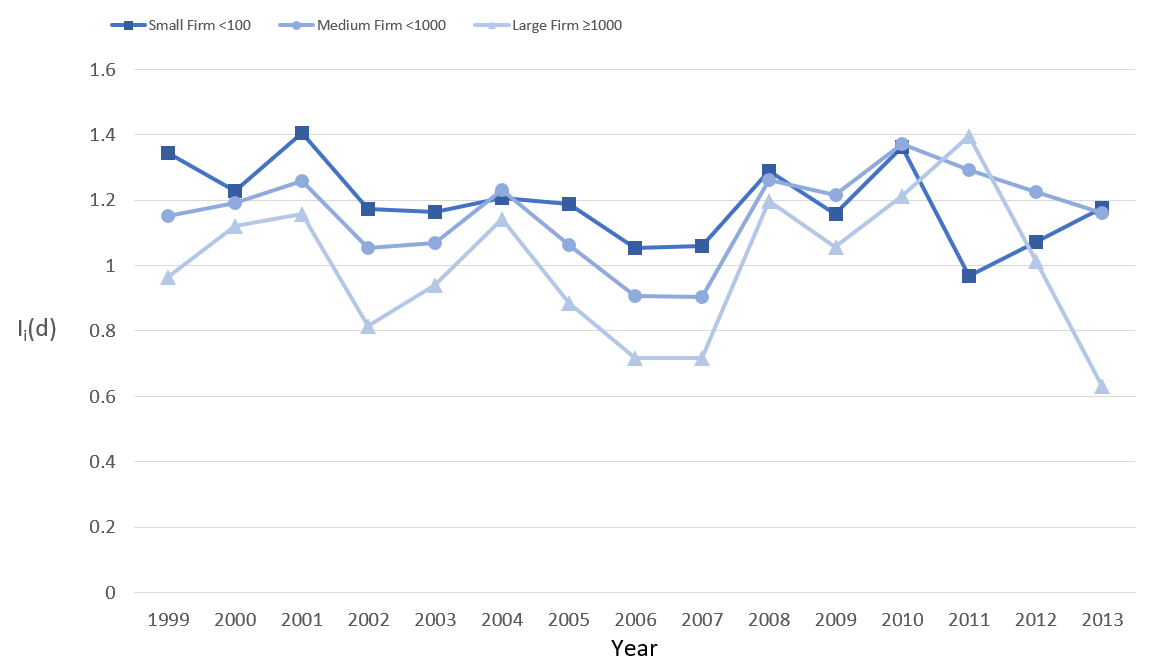}
\par\end{center}

\noindent \begin{center}
Figure 2 The transition entropy for small, medium and large firms
from 1999 to 2013
\par\end{center}

\noindent \begin{center}
{\tiny{}Year in this figure means end year. $I_{i}(d)$ is transition
entropy. Full data see Appendix.}{\tiny\par}
\par\end{center}

\subsection{Robustness check of Markov transfer matrices}

According to Neumark et al. (2011), the obtained results should be
checked by 2-year periods instead of 1 in the consideration of robustness.
To this end, the 2nd-ordered Markov transfer matrices will be calculated
and compared with the previous results. According to equation (7),
when $P(0)$ is given,

\begin{equation}
P(d)=F(d-1,d)F(d-2,d-1)P(d-2)=F(d-2,d)P(d-2),
\end{equation}

where $d\geq2$ and $F(d-2,d)$ is the second-ordered Markov transfer
matrix from $d-2$ to $d$. This means the second-ordered Markov transfer
matrices should completely equals to the product of two continuous
first-order Markov transfer matrices. Table 3 is the second-order
transfer matrix 1998-2000 calculated from the product of two first-order
transfer matrices 1998-1999 and 1999-2000. Table 4 is the second-order
transfer matrix 1998-2000 calculated by the cross-sectional data from
1998 to 2000.

Note that the later cannot be calculated by only the two cross-sectional
data sets of 1998 and 2000. This is because the firms that enter at
1999 while exit at 2000 are ignored, which are considered as 0-to-0
state transfers from 1998 to 2000. To include all 0-to-0 state transfer
in the second-order matrices, the three cross-sectional data sets
of 1998, 1999 and 2000 must be rectangularized to a strongly balanced
data set with missing values replaced by 0. With this change, all
firms that exist in any of the three years are included, and it can
be found that the second-order transfer matrix obtained from data
set equals to the product of two first-order transfer matrix, which
means the obtained first-order transfer matrices are correctly calculated.

Compare the second-order transfer matrix with the first-order transfer
matrix, it can be found that there are no significant differences
between most of the elements. The most obvious difference between
the two kinds is that compared to the second-order matrices, the diagonal
elements of the first-order matrices have larger values. This can
be understood as when time interval length increases, the probability
that firms remain in the same categories may gradually decrease. But
in general, it can be considered that the results obtained are not
very sensitive to the time interval length.
\noindent \begin{center}
Table 3 The second-order transfer matrix 1998-2000 (calculated from
product)
\par\end{center}

\noindent \begin{center}
{\tiny{}}%
\begin{tabular}{llllllllllllll}
\hline 
{\tiny{}Size} & {\tiny{}0} & {\tiny{}1} & {\tiny{}2} & {\tiny{}3} & {\tiny{}4} & {\tiny{}5} & {\tiny{}6} & {\tiny{}7} & {\tiny{}8} & {\tiny{}9} & {\tiny{}10} & {\tiny{}11} & {\tiny{}12}\tabularnewline
\hline 
{\tiny{}0} & {\tiny{}0.2165} & {\tiny{}0.2074} & {\tiny{}0.1968} & {\tiny{}0.1787} & {\tiny{}0.1588} & {\tiny{}0.1427} & {\tiny{}0.1319} & {\tiny{}0.1290} & {\tiny{}0.1261} & {\tiny{}0.1354} & {\tiny{}0.1441} & {\tiny{}0.1143} & {\tiny{}0.1987}\tabularnewline
{\tiny{}1} & {\tiny{}0.0495} & {\tiny{}0.2219} & {\tiny{}0.0834} & {\tiny{}0.0485} & {\tiny{}0.0356} & {\tiny{}0.0261} & {\tiny{}0.0201} & {\tiny{}0.0177} & {\tiny{}0.0118} & {\tiny{}0.0111} & {\tiny{}0.0081} & {\tiny{}0.0088} & {\tiny{}0.0027}\tabularnewline
{\tiny{}2} & {\tiny{}0.1251} & {\tiny{}0.1814} & {\tiny{}0.3274} & {\tiny{}0.1408} & {\tiny{}0.0661} & {\tiny{}0.0427} & {\tiny{}0.0335} & {\tiny{}0.0296} & {\tiny{}0.0246} & {\tiny{}0.0216} & {\tiny{}0.0173} & {\tiny{}0.0193} & {\tiny{}0.0060}\tabularnewline
{\tiny{}3} & {\tiny{}0.1789} & {\tiny{}0.1488} & {\tiny{}0.1931} & {\tiny{}0.3704} & {\tiny{}0.1487} & {\tiny{}0.0678} & {\tiny{}0.0470} & {\tiny{}0.0378} & {\tiny{}0.0308} & {\tiny{}0.0282} & {\tiny{}0.0265} & {\tiny{}0.0248} & {\tiny{}0.0074}\tabularnewline
{\tiny{}4} & {\tiny{}0.2345} & {\tiny{}0.1532} & {\tiny{}0.1298} & {\tiny{}0.1970} & {\tiny{}0.4676} & {\tiny{}0.2134} & {\tiny{}0.0929} & {\tiny{}0.0579} & {\tiny{}0.0388} & {\tiny{}0.0324} & {\tiny{}0.0258} & {\tiny{}0.0262} & {\tiny{}0.0080}\tabularnewline
{\tiny{}5} & {\tiny{}0.1076} & {\tiny{}0.0542} & {\tiny{}0.0451} & {\tiny{}0.0428} & {\tiny{}0.0971} & {\tiny{}0.4145} & {\tiny{}0.1870} & {\tiny{}0.0584} & {\tiny{}0.0314} & {\tiny{}0.0200} & {\tiny{}0.0145} & {\tiny{}0.0116} & {\tiny{}0.0035}\tabularnewline
{\tiny{}6} & {\tiny{}0.0527} & {\tiny{}0.0192} & {\tiny{}0.0158} & {\tiny{}0.0149} & {\tiny{}0.0187} & {\tiny{}0.0795} & {\tiny{}0.4210} & {\tiny{}0.1730} & {\tiny{}0.0484} & {\tiny{}0.0196} & {\tiny{}0.0141} & {\tiny{}0.0067} & {\tiny{}0.0017}\tabularnewline
{\tiny{}7} & {\tiny{}0.0258} & {\tiny{}0.0115} & {\tiny{}0.0067} & {\tiny{}0.0056} & {\tiny{}0.0058} & {\tiny{}0.0109} & {\tiny{}0.0616} & {\tiny{}0.4567} & {\tiny{}0.2071} & {\tiny{}0.0493} & {\tiny{}0.0259} & {\tiny{}0.0208} & {\tiny{}0.0086}\tabularnewline
{\tiny{}8} & {\tiny{}0.0057} & {\tiny{}0.0015} & {\tiny{}0.0011} & {\tiny{}0.0009} & {\tiny{}0.0009} & {\tiny{}0.0018} & {\tiny{}0.0037} & {\tiny{}0.0358} & {\tiny{}0.4271} & {\tiny{}0.1906} & {\tiny{}0.0335} & {\tiny{}0.0155} & {\tiny{}0.0181}\tabularnewline
{\tiny{}9} & {\tiny{}0.0023} & {\tiny{}0.0005} & {\tiny{}0.0004} & {\tiny{}0.0004} & {\tiny{}0.0002} & {\tiny{}0.0004} & {\tiny{}0.0008} & {\tiny{}0.0032} & {\tiny{}0.0472} & {\tiny{}0.4547} & {\tiny{}0.2135} & {\tiny{}0.0335} & {\tiny{}0.0012}\tabularnewline
{\tiny{}10} & {\tiny{}0.0009} & {\tiny{}0.0003} & {\tiny{}0.0002} & {\tiny{}0.0002} & {\tiny{}0.0001} & {\tiny{}0.0001} & {\tiny{}0.0003} & {\tiny{}0.0009} & {\tiny{}0.0062} & {\tiny{}0.0365} & {\tiny{}0.4479} & {\tiny{}0.2684} & {\tiny{}0.0739}\tabularnewline
{\tiny{}11} & {\tiny{}0.0002} & {\tiny{}0.0000} & {\tiny{}0.0000} & {\tiny{}0.0000} & {\tiny{}0.0000} & {\tiny{}0.0000} & {\tiny{}0.0002} & {\tiny{}0.0000} & {\tiny{}0.0005} & {\tiny{}0.0005} & {\tiny{}0.0260} & {\tiny{}0.3548} & {\tiny{}0.1596}\tabularnewline
{\tiny{}12} & {\tiny{}0.0002} & {\tiny{}0.0001} & {\tiny{}0.0001} & {\tiny{}0.0000} & {\tiny{}0.0000} & {\tiny{}0.0000} & {\tiny{}0.0001} & {\tiny{}0.0000} & {\tiny{}0.0000} & {\tiny{}0.0000} & {\tiny{}0.0027} & {\tiny{}0.0952} & {\tiny{}0.5106}\tabularnewline
\hline 
\multicolumn{14}{r}{{\tiny{}Axis coordinates represent size categories, transfer probabilities
in the matrix.}}\tabularnewline
\end{tabular}{\tiny\par}
\par\end{center}

\pagebreak{}
\noindent \begin{center}
Table 4 The second-order transfer matrix 1998-2000 (calculated from
data)
\par\end{center}

\noindent \begin{center}
{\tiny{}}%
\begin{tabular}{llllllllllllll}
\hline 
{\tiny{}Size} & {\tiny{}0} & {\tiny{}1} & {\tiny{}2} & {\tiny{}3} & {\tiny{}4} & {\tiny{}5} & {\tiny{}6} & {\tiny{}7} & {\tiny{}8} & {\tiny{}9} & {\tiny{}10} & {\tiny{}11} & {\tiny{}12}\tabularnewline
\hline 
{\tiny{}0} & {\tiny{}0.2166} & {\tiny{}0.2074} & {\tiny{}0.1968} & {\tiny{}0.1786} & {\tiny{}0.1588} & {\tiny{}0.1426} & {\tiny{}0.1318} & {\tiny{}0.1290} & {\tiny{}0.1260} & {\tiny{}0.1354} & {\tiny{}0.1442} & {\tiny{}0.1144} & {\tiny{}0.1987}\tabularnewline
{\tiny{}1} & {\tiny{}0.0496} & {\tiny{}0.2219} & {\tiny{}0.0835} & {\tiny{}0.0485} & {\tiny{}0.0356} & {\tiny{}0.0262} & {\tiny{}0.0200} & {\tiny{}0.0177} & {\tiny{}0.0118} & {\tiny{}0.0111} & {\tiny{}0.0081} & {\tiny{}0.0088} & {\tiny{}0.0027}\tabularnewline
{\tiny{}2} & {\tiny{}0.1251} & {\tiny{}0.1813} & {\tiny{}0.3274} & {\tiny{}0.1409} & {\tiny{}0.0660} & {\tiny{}0.0426} & {\tiny{}0.0335} & {\tiny{}0.0296} & {\tiny{}0.0246} & {\tiny{}0.0217} & {\tiny{}0.0173} & {\tiny{}0.0193} & {\tiny{}0.0061}\tabularnewline
{\tiny{}3} & {\tiny{}0.1789} & {\tiny{}0.1488} & {\tiny{}0.1931} & {\tiny{}0.3705} & {\tiny{}0.1488} & {\tiny{}0.0679} & {\tiny{}0.0471} & {\tiny{}0.0378} & {\tiny{}0.0309} & {\tiny{}0.0282} & {\tiny{}0.0265} & {\tiny{}0.0248} & {\tiny{}0.0074}\tabularnewline
{\tiny{}4} & {\tiny{}0.2346} & {\tiny{}0.1532} & {\tiny{}0.1298} & {\tiny{}0.1970} & {\tiny{}0.4675} & {\tiny{}0.2134} & {\tiny{}0.0929} & {\tiny{}0.0580} & {\tiny{}0.0388} & {\tiny{}0.0324} & {\tiny{}0.0259} & {\tiny{}0.0263} & {\tiny{}0.0081}\tabularnewline
{\tiny{}5} & {\tiny{}0.1076} & {\tiny{}0.0542} & {\tiny{}0.0452} & {\tiny{}0.0427} & {\tiny{}0.0970} & {\tiny{}0.4144} & {\tiny{}0.1871} & {\tiny{}0.0583} & {\tiny{}0.0315} & {\tiny{}0.0200} & {\tiny{}0.0145} & {\tiny{}0.0115} & {\tiny{}0.0035}\tabularnewline
{\tiny{}6} & {\tiny{}0.0527} & {\tiny{}0.0192} & {\tiny{}0.0157} & {\tiny{}0.0150} & {\tiny{}0.0187} & {\tiny{}0.0795} & {\tiny{}0.4210} & {\tiny{}0.1729} & {\tiny{}0.0484} & {\tiny{}0.0196} & {\tiny{}0.0141} & {\tiny{}0.0068} & {\tiny{}0.0016}\tabularnewline
{\tiny{}7} & {\tiny{}0.0259} & {\tiny{}0.0115} & {\tiny{}0.0066} & {\tiny{}0.0057} & {\tiny{}0.0058} & {\tiny{}0.0109} & {\tiny{}0.0616} & {\tiny{}0.4566} & {\tiny{}0.2071} & {\tiny{}0.0492} & {\tiny{}0.0259} & {\tiny{}0.0209} & {\tiny{}0.0086}\tabularnewline
{\tiny{}8} & {\tiny{}0.0057} & {\tiny{}0.0016} & {\tiny{}0.0012} & {\tiny{}0.0008} & {\tiny{}0.0009} & {\tiny{}0.0018} & {\tiny{}0.0037} & {\tiny{}0.0358} & {\tiny{}0.4271} & {\tiny{}0.1906} & {\tiny{}0.0336} & {\tiny{}0.0156} & {\tiny{}0.0181}\tabularnewline
{\tiny{}9} & {\tiny{}0.0022} & {\tiny{}0.0004} & {\tiny{}0.0004} & {\tiny{}0.0005} & {\tiny{}0.0002} & {\tiny{}0.0004} & {\tiny{}0.0008} & {\tiny{}0.0033} & {\tiny{}0.0472} & {\tiny{}0.4546} & {\tiny{}0.2136} & {\tiny{}0.0336} & {\tiny{}0.0013}\tabularnewline
{\tiny{}10} & {\tiny{}0.0009} & {\tiny{}0.0002} & {\tiny{}0.0002} & {\tiny{}0.0002} & {\tiny{}0.0001} & {\tiny{}0.0002} & {\tiny{}0.0004} & {\tiny{}0.0008} & {\tiny{}0.0063} & {\tiny{}0.0366} & {\tiny{}0.4479} & {\tiny{}0.2685} & {\tiny{}0.0740}\tabularnewline
{\tiny{}11} & {\tiny{}0.0002} & {\tiny{}0.0000} & {\tiny{}0.0000} & {\tiny{}0.0000} & {\tiny{}0.0000} & {\tiny{}0.0000} & {\tiny{}0.0002} & {\tiny{}0.0000} & {\tiny{}0.0004} & {\tiny{}0.0004} & {\tiny{}0.0260} & {\tiny{}0.3548} & {\tiny{}0.1596}\tabularnewline
{\tiny{}12} & {\tiny{}0.0002} & {\tiny{}0.0001} & {\tiny{}0.0001} & {\tiny{}0.0000} & {\tiny{}0.0000} & {\tiny{}0.0000} & {\tiny{}0.0001} & {\tiny{}0.0000} & {\tiny{}0.0001} & {\tiny{}0.0000} & {\tiny{}0.0027} & {\tiny{}0.0951} & {\tiny{}0.5106}\tabularnewline
\hline 
\multicolumn{14}{r}{{\tiny{}Axis coordinates represent size categories, transfer probabilities
in the matrix.}}\tabularnewline
\end{tabular}{\tiny\par}
\par\end{center}

\section{Conclusion and Discussion}

In this paper, a Markov-chain-based descriptive method is proposed
for presenting firm size transition dynamics. By dividing firms into
multiple states according to their size categories, a Markov chain
that reveals the internal transfer between size categories is established.
Using the properties of Markov chain, the definition of transition
path, transition trend and transition entropy are introduced based
on first-order transition matrices. By calculating the transition
path, the growth probabilities of firms in terms of size for a certain
time period can be clearly demonstrated; by evaluating the transition
trend, the tendency of firm size transition towards large firms or
small and medium firms in any time period can be revealed; by calculating
transition entropy of a certain firm size category, the uncertainty
of firm size transition, or in other words, the likelihood of firm
size change can be quantified.

Furthermore, the proposed descriptive method is used on the rich firm-level
data set extracted from the Chinese Industrial Enterprises Database
1998-2013. From the transition matrices it can be observed that most
firms have minor size changes, while a small number of firms may grow
or deteriorate significantly. From the result of transition trend,
it can be found that in 2004, 2010 and 2012 Chinese manufacturing
firms tend to transfer into larger firms, which implies economic growth;
in 2008, 2009 and 2013 however, firms tend to transfer into smaller
firms, which implies economic recession. From the result of transition
entropy, it can be observed that compared to large firms, small and
medium firms in China are much more likely to change in terms of size.
In general, evidence concludes that small and medium manufacturing
firms in China have greater job creation potentials compared to large
firms over the time period.

Nevertheless, the research remains imperfect. First, the proposed
Markov-chain-based method is only applicable to describe the fact
that has occurred. There is no function of prediction so far. In addition,
currently there is no sufficient precedent paper fully investigate
job creation issue via Markov chain. So, whether the property of Markov
chains is applicable to empirical analysis and forecasting under all
circumstances needs further discussion. Finally, this paper follows
the classification standard proposed by Neumark et al. (2011) to classify
firm size, which is extensively used but not linear. According to
previous research, it can be roughly considered that this classification
conforms to a log-normal function (You \& Fan, 2020), yet this standard
might not be the best on describing categories of small firms.

\pagebreak{}

\part*{Appendix}

\subsection*{Obtained First-order Markov Transfer Matrices}

The obtained first-order Markov transfer matrices from 1998 to 2013
are presented as follows:
\noindent \begin{center}
Table 5 The first-order transfer matrix from 1998 to 1999
\par\end{center}

\noindent \begin{center}
{\tiny{}}%
\begin{tabular}{llllllllllllll}
\hline 
{\tiny{}Size} & {\tiny{}0} & {\tiny{}1} & {\tiny{}2} & {\tiny{}3} & {\tiny{}4} & {\tiny{}5} & {\tiny{}6} & {\tiny{}7} & {\tiny{}8} & {\tiny{}9} & {\tiny{}10} & {\tiny{}11} & {\tiny{}12}\tabularnewline
\hline 
{\tiny{}0} & {\tiny{}0.0000} & {\tiny{}0.3688} & {\tiny{}0.2768} & {\tiny{}0.2173} & {\tiny{}0.1743} & {\tiny{}0.1366} & {\tiny{}0.1186} & {\tiny{}0.1219} & {\tiny{}0.1061} & {\tiny{}0.1024} & {\tiny{}0.0815} & {\tiny{}0.0926} & {\tiny{}0.0286}\tabularnewline
{\tiny{}1} & {\tiny{}0.0546} & {\tiny{}0.3947} & {\tiny{}0.0526} & {\tiny{}0.0122} & {\tiny{}0.0055} & {\tiny{}0.0029} & {\tiny{}0.0014} & {\tiny{}0.0017} & {\tiny{}0.0000} & {\tiny{}0.0000} & {\tiny{}0.0000} & {\tiny{}0.0000} & {\tiny{}0.0000}\tabularnewline
{\tiny{}2} & {\tiny{}0.1759} & {\tiny{}0.1099} & {\tiny{}0.4844} & {\tiny{}0.0914} & {\tiny{}0.0163} & {\tiny{}0.0068} & {\tiny{}0.0052} & {\tiny{}0.0018} & {\tiny{}0.0019} & {\tiny{}0.0000} & {\tiny{}0.0000} & {\tiny{}0.0000} & {\tiny{}0.0000}\tabularnewline
{\tiny{}3} & {\tiny{}0.2425} & {\tiny{}0.0455} & {\tiny{}0.1223} & {\tiny{}0.5370} & {\tiny{}0.0921} & {\tiny{}0.0165} & {\tiny{}0.0077} & {\tiny{}0.0024} & {\tiny{}0.0019} & {\tiny{}0.0000} & {\tiny{}0.0043} & {\tiny{}0.0000} & {\tiny{}0.0000}\tabularnewline
{\tiny{}4} & {\tiny{}0.2952} & {\tiny{}0.0566} & {\tiny{}0.0421} & {\tiny{}0.1235} & {\tiny{}0.6403} & {\tiny{}0.1501} & {\tiny{}0.0336} & {\tiny{}0.0091} & {\tiny{}0.0025} & {\tiny{}0.0017} & {\tiny{}0.0000} & {\tiny{}0.0000} & {\tiny{}0.0000}\tabularnewline
{\tiny{}5} & {\tiny{}0.1279} & {\tiny{}0.0160} & {\tiny{}0.0168} & {\tiny{}0.0121} & {\tiny{}0.0628} & {\tiny{}0.6264} & {\tiny{}0.1542} & {\tiny{}0.0254} & {\tiny{}0.0025} & {\tiny{}0.0068} & {\tiny{}0.0000} & {\tiny{}0.0000} & {\tiny{}0.0000}\tabularnewline
\hline 
\end{tabular}{\tiny\par}
\par\end{center}

\noindent \begin{center}
Table 6 The first-order transfer matrix from 1999 to 2000
\par\end{center}

\noindent \begin{center}
{\tiny{}}%
\begin{tabular}{llllllllllllll}
\hline 
{\tiny{}Size} & {\tiny{}0} & {\tiny{}1} & {\tiny{}2} & {\tiny{}3} & {\tiny{}4} & {\tiny{}5} & {\tiny{}6} & {\tiny{}7} & {\tiny{}8} & {\tiny{}9} & {\tiny{}10} & {\tiny{}11} & {\tiny{}12}\tabularnewline
\hline 
{\tiny{}0} & {\tiny{}0.0000} & {\tiny{}0.3837} & {\tiny{}0.2840} & {\tiny{}0.2272} & {\tiny{}0.1874} & {\tiny{}0.1582} & {\tiny{}0.1433} & {\tiny{}0.1457} & {\tiny{}0.1382} & {\tiny{}0.1538} & {\tiny{}0.1587} & {\tiny{}0.1111} & {\tiny{}0.2424}\tabularnewline
{\tiny{}1} & {\tiny{}0.0951} & {\tiny{}0.4501} & {\tiny{}0.0601} & {\tiny{}0.0266} & {\tiny{}0.0192} & {\tiny{}0.0121} & {\tiny{}0.0077} & {\tiny{}0.0054} & {\tiny{}0.0006} & {\tiny{}0.0016} & {\tiny{}0.0000} & {\tiny{}0.0000} & {\tiny{}0.0000}\tabularnewline
{\tiny{}2} & {\tiny{}0.2079} & {\tiny{}0.1071} & {\tiny{}0.5227} & {\tiny{}0.0819} & {\tiny{}0.0197} & {\tiny{}0.0093} & {\tiny{}0.0050} & {\tiny{}0.0026} & {\tiny{}0.0012} & {\tiny{}0.0000} & {\tiny{}0.0000} & {\tiny{}0.0000} & {\tiny{}0.0000}\tabularnewline
{\tiny{}3} & {\tiny{}0.2575} & {\tiny{}0.0326} & {\tiny{}0.1018} & {\tiny{}0.5490} & {\tiny{}0.0786} & {\tiny{}0.0166} & {\tiny{}0.0100} & {\tiny{}0.0034} & {\tiny{}0.0018} & {\tiny{}0.0016} & {\tiny{}0.0040} & {\tiny{}0.0000} & {\tiny{}0.0000}\tabularnewline
{\tiny{}4} & {\tiny{}0.2718} & {\tiny{}0.0196} & {\tiny{}0.0250} & {\tiny{}0.1046} & {\tiny{}0.6278} & {\tiny{}0.1252} & {\tiny{}0.0292} & {\tiny{}0.0168} & {\tiny{}0.0066} & {\tiny{}0.0016} & {\tiny{}0.0040} & {\tiny{}0.0000} & {\tiny{}0.0000}\tabularnewline
{\tiny{}5} & {\tiny{}0.1026} & {\tiny{}0.0046} & {\tiny{}0.0044} & {\tiny{}0.0083} & {\tiny{}0.0608} & {\tiny{}0.6136} & {\tiny{}0.1218} & {\tiny{}0.0181} & {\tiny{}0.0203} & {\tiny{}0.0016} & {\tiny{}0.0079} & {\tiny{}0.0000} & {\tiny{}0.0000}\tabularnewline
\hline 
\end{tabular}{\tiny\par}
\par\end{center}

\noindent \begin{center}
Table 7 The first-order transfer matrix from 2000 to 2001
\par\end{center}

\noindent \begin{center}
{\tiny{}}%
\begin{tabular}{llllllllllllll}
\hline 
{\tiny{}Size} & {\tiny{}0} & {\tiny{}1} & {\tiny{}2} & {\tiny{}3} & {\tiny{}4} & {\tiny{}5} & {\tiny{}6} & {\tiny{}7} & {\tiny{}8} & {\tiny{}9} & {\tiny{}10} & {\tiny{}11} & {\tiny{}12}\tabularnewline
\hline 
{\tiny{}0} & {\tiny{}0.0000} & {\tiny{}0.4047} & {\tiny{}0.3289} & {\tiny{}0.2718} & {\tiny{}0.2386} & {\tiny{}0.2143} & {\tiny{}0.1967} & {\tiny{}0.1857} & {\tiny{}0.1828} & {\tiny{}0.1813} & {\tiny{}0.1187} & {\tiny{}0.0465} & {\tiny{}0.1795}\tabularnewline
{\tiny{}1} & {\tiny{}0.0586} & {\tiny{}0.2635} & {\tiny{}0.0502} & {\tiny{}0.0115} & {\tiny{}0.0056} & {\tiny{}0.0037} & {\tiny{}0.0023} & {\tiny{}0.0018} & {\tiny{}0.0000} & {\tiny{}0.0000} & {\tiny{}0.0000} & {\tiny{}0.0000} & {\tiny{}0.0000}\tabularnewline
{\tiny{}2} & {\tiny{}0.1947} & {\tiny{}0.1276} & {\tiny{}0.4332} & {\tiny{}0.0879} & {\tiny{}0.0184} & {\tiny{}0.0074} & {\tiny{}0.0037} & {\tiny{}0.0018} & {\tiny{}0.0000} & {\tiny{}0.0000} & {\tiny{}0.0000} & {\tiny{}0.0000} & {\tiny{}0.0000}\tabularnewline
{\tiny{}3} & {\tiny{}0.2659} & {\tiny{}0.0841} & {\tiny{}0.1323} & {\tiny{}0.4812} & {\tiny{}0.0925} & {\tiny{}0.0158} & {\tiny{}0.0061} & {\tiny{}0.0030} & {\tiny{}0.0020} & {\tiny{}0.0000} & {\tiny{}0.0000} & {\tiny{}0.0000} & {\tiny{}0.0000}\tabularnewline
{\tiny{}4} & {\tiny{}0.2891} & {\tiny{}0.0804} & {\tiny{}0.0441} & {\tiny{}0.1292} & {\tiny{}0.5658} & {\tiny{}0.1551} & {\tiny{}0.0332} & {\tiny{}0.0119} & {\tiny{}0.0055} & {\tiny{}0.0000} & {\tiny{}0.0046} & {\tiny{}0.0233} & {\tiny{}0.0000}\tabularnewline
{\tiny{}5} & {\tiny{}0.1143} & {\tiny{}0.0249} & {\tiny{}0.0092} & {\tiny{}0.0143} & {\tiny{}0.0690} & {\tiny{}0.5387} & {\tiny{}0.1558} & {\tiny{}0.0242} & {\tiny{}0.0130} & {\tiny{}0.0072} & {\tiny{}0.0000} & {\tiny{}0.0000} & {\tiny{}0.0000}\tabularnewline
\hline 
\end{tabular}{\tiny\par}
\par\end{center}

\pagebreak{}
\noindent \begin{center}
Table 8 The first-order transfer matrix from 2001 to 2002
\par\end{center}

\noindent \begin{center}
{\tiny{}}%
\begin{tabular}{llllllllllllll}
\hline 
{\tiny{}Size} & {\tiny{}0} & {\tiny{}1} & {\tiny{}2} & {\tiny{}3} & {\tiny{}4} & {\tiny{}5} & {\tiny{}6} & {\tiny{}7} & {\tiny{}8} & {\tiny{}9} & {\tiny{}10} & {\tiny{}11} & {\tiny{}12}\tabularnewline
\hline 
{\tiny{}0} & {\tiny{}0.0000} & {\tiny{}0.3372} & {\tiny{}0.2247} & {\tiny{}0.1651} & {\tiny{}0.1272} & {\tiny{}0.1014} & {\tiny{}0.0869} & {\tiny{}0.0744} & {\tiny{}0.0724} & {\tiny{}0.0633} & {\tiny{}0.0842} & {\tiny{}0.0800} & {\tiny{}0.0000}\tabularnewline
{\tiny{}1} & {\tiny{}0.0548} & {\tiny{}0.4986} & {\tiny{}0.0503} & {\tiny{}0.0110} & {\tiny{}0.0053} & {\tiny{}0.0029} & {\tiny{}0.0020} & {\tiny{}0.0023} & {\tiny{}0.0000} & {\tiny{}0.0000} & {\tiny{}0.0000} & {\tiny{}0.0000} & {\tiny{}0.0000}\tabularnewline
{\tiny{}2} & {\tiny{}0.2198} & {\tiny{}0.1140} & {\tiny{}0.5684} & {\tiny{}0.0825} & {\tiny{}0.0164} & {\tiny{}0.0064} & {\tiny{}0.0029} & {\tiny{}0.0024} & {\tiny{}0.0000} & {\tiny{}0.0000} & {\tiny{}0.0000} & {\tiny{}0.0000} & {\tiny{}0.0000}\tabularnewline
{\tiny{}3} & {\tiny{}0.2800} & {\tiny{}0.0257} & {\tiny{}0.1239} & {\tiny{}0.6103} & {\tiny{}0.0842} & {\tiny{}0.0129} & {\tiny{}0.0062} & {\tiny{}0.0016} & {\tiny{}0.0007} & {\tiny{}0.0000} & {\tiny{}0.0000} & {\tiny{}0.0000} & {\tiny{}0.0000}\tabularnewline
{\tiny{}4} & {\tiny{}0.2853} & {\tiny{}0.0194} & {\tiny{}0.0272} & {\tiny{}0.1208} & {\tiny{}0.6937} & {\tiny{}0.1354} & {\tiny{}0.0222} & {\tiny{}0.0104} & {\tiny{}0.0043} & {\tiny{}0.0037} & {\tiny{}0.0105} & {\tiny{}0.0000} & {\tiny{}0.0385}\tabularnewline
{\tiny{}5} & {\tiny{}0.0960} & {\tiny{}0.0043} & {\tiny{}0.0047} & {\tiny{}0.0082} & {\tiny{}0.0659} & {\tiny{}0.6721} & {\tiny{}0.1329} & {\tiny{}0.0163} & {\tiny{}0.0135} & {\tiny{}0.0074} & {\tiny{}0.0000} & {\tiny{}0.0000} & {\tiny{}0.0000}\tabularnewline
\hline 
\end{tabular}{\tiny\par}
\par\end{center}

\noindent \begin{center}
Table 9 The first-order transfer matrix from 2002 to 2003
\par\end{center}

\noindent \begin{center}
{\tiny{}}%
\begin{tabular}{llllllllllllll}
\hline 
{\tiny{}Size} & {\tiny{}0} & {\tiny{}1} & {\tiny{}2} & {\tiny{}3} & {\tiny{}4} & {\tiny{}5} & {\tiny{}6} & {\tiny{}7} & {\tiny{}8} & {\tiny{}9} & {\tiny{}10} & {\tiny{}11} & {\tiny{}12}\tabularnewline
\hline 
{\tiny{}0} & {\tiny{}0.0000} & {\tiny{}0.4219} & {\tiny{}0.2423} & {\tiny{}0.1800} & {\tiny{}0.1440} & {\tiny{}0.1164} & {\tiny{}0.1014} & {\tiny{}0.0919} & {\tiny{}0.1042} & {\tiny{}0.1612} & {\tiny{}0.0821} & {\tiny{}0.0652} & {\tiny{}0.1111}\tabularnewline
{\tiny{}1} & {\tiny{}0.0469} & {\tiny{}0.4257} & {\tiny{}0.0398} & {\tiny{}0.0084} & {\tiny{}0.0035} & {\tiny{}0.0024} & {\tiny{}0.0019} & {\tiny{}0.0008} & {\tiny{}0.0013} & {\tiny{}0.0000} & {\tiny{}0.0000} & {\tiny{}0.0000} & {\tiny{}0.0000}\tabularnewline
{\tiny{}2} & {\tiny{}0.2315} & {\tiny{}0.1105} & {\tiny{}0.5546} & {\tiny{}0.0754} & {\tiny{}0.0133} & {\tiny{}0.0044} & {\tiny{}0.0016} & {\tiny{}0.0011} & {\tiny{}0.0007} & {\tiny{}0.0018} & {\tiny{}0.0000} & {\tiny{}0.0000} & {\tiny{}0.0000}\tabularnewline
{\tiny{}3} & {\tiny{}0.2947} & {\tiny{}0.0255} & {\tiny{}0.1325} & {\tiny{}0.6030} & {\tiny{}0.0803} & {\tiny{}0.0108} & {\tiny{}0.0061} & {\tiny{}0.0026} & {\tiny{}0.0007} & {\tiny{}0.0036} & {\tiny{}0.0000} & {\tiny{}0.0000} & {\tiny{}0.0000}\tabularnewline
{\tiny{}4} & {\tiny{}0.2796} & {\tiny{}0.0125} & {\tiny{}0.0259} & {\tiny{}0.1230} & {\tiny{}0.6794} & {\tiny{}0.1220} & {\tiny{}0.0198} & {\tiny{}0.0099} & {\tiny{}0.0046} & {\tiny{}0.0036} & {\tiny{}0.0051} & {\tiny{}0.0000} & {\tiny{}0.0000}\tabularnewline
{\tiny{}5} & {\tiny{}0.0933} & {\tiny{}0.0022} & {\tiny{}0.0037} & {\tiny{}0.0078} & {\tiny{}0.0724} & {\tiny{}0.6620} & {\tiny{}0.1222} & {\tiny{}0.0124} & {\tiny{}0.0073} & {\tiny{}0.0054} & {\tiny{}0.0051} & {\tiny{}0.0000} & {\tiny{}0.0000}\tabularnewline
\hline 
\end{tabular}{\tiny\par}
\par\end{center}

\noindent \begin{center}
Table 10 The first-order transfer matrix from 2003 to 2004
\par\end{center}

\noindent \begin{center}
{\tiny{}}%
\begin{tabular}{llllllllllllll}
\hline 
{\tiny{}Size} & {\tiny{}0} & {\tiny{}1} & {\tiny{}2} & {\tiny{}3} & {\tiny{}4} & {\tiny{}5} & {\tiny{}6} & {\tiny{}7} & {\tiny{}8} & {\tiny{}9} & {\tiny{}10} & {\tiny{}11} & {\tiny{}12}\tabularnewline
\hline 
{\tiny{}0} & {\tiny{}0.0000} & {\tiny{}0.5122} & {\tiny{}0.3566} & {\tiny{}0.2821} & {\tiny{}0.2334} & {\tiny{}0.1984} & {\tiny{}0.1758} & {\tiny{}0.1692} & {\tiny{}0.1691} & {\tiny{}0.1715} & {\tiny{}0.1611} & {\tiny{}0.1207} & {\tiny{}0.0400}\tabularnewline
{\tiny{}1} & {\tiny{}0.0808} & {\tiny{}0.3336} & {\tiny{}0.0361} & {\tiny{}0.0062} & {\tiny{}0.0034} & {\tiny{}0.0022} & {\tiny{}0.0016} & {\tiny{}0.0008} & {\tiny{}0.0006} & {\tiny{}0.0019} & {\tiny{}0.0000} & {\tiny{}0.0000} & {\tiny{}0.0000}\tabularnewline
{\tiny{}2} & {\tiny{}0.2631} & {\tiny{}0.1118} & {\tiny{}0.4433} & {\tiny{}0.0771} & {\tiny{}0.0156} & {\tiny{}0.0070} & {\tiny{}0.0036} & {\tiny{}0.0034} & {\tiny{}0.0057} & {\tiny{}0.0000} & {\tiny{}0.0000} & {\tiny{}0.0000} & {\tiny{}0.0000}\tabularnewline
{\tiny{}3} & {\tiny{}0.2893} & {\tiny{}0.0234} & {\tiny{}0.1294} & {\tiny{}0.5001} & {\tiny{}0.0859} & {\tiny{}0.0161} & {\tiny{}0.0086} & {\tiny{}0.0035} & {\tiny{}0.0051} & {\tiny{}0.0039} & {\tiny{}0.0000} & {\tiny{}0.0000} & {\tiny{}0.0000}\tabularnewline
{\tiny{}4} & {\tiny{}0.2469} & {\tiny{}0.0142} & {\tiny{}0.0300} & {\tiny{}0.1237} & {\tiny{}0.5819} & {\tiny{}0.1376} & {\tiny{}0.0310} & {\tiny{}0.0143} & {\tiny{}0.0102} & {\tiny{}0.0039} & {\tiny{}0.0000} & {\tiny{}0.0000} & {\tiny{}0.0000}\tabularnewline
{\tiny{}5} & {\tiny{}0.0776} & {\tiny{}0.0018} & {\tiny{}0.0039} & {\tiny{}0.0093} & {\tiny{}0.0725} & {\tiny{}0.5615} & {\tiny{}0.1318} & {\tiny{}0.0247} & {\tiny{}0.0153} & {\tiny{}0.0058} & {\tiny{}0.0111} & {\tiny{}0.0000} & {\tiny{}0.0000}\tabularnewline
\hline 
\end{tabular}{\tiny\par}
\par\end{center}

\pagebreak{}
\noindent \begin{center}
Table 11 The first-order transfer matrix from 2004 to 2005
\par\end{center}

\noindent \begin{center}
{\tiny{}}%
\begin{tabular}{llllllllllllll}
\hline 
{\tiny{}Size} & {\tiny{}0} & {\tiny{}1} & {\tiny{}2} & {\tiny{}3} & {\tiny{}4} & {\tiny{}5} & {\tiny{}6} & {\tiny{}7} & {\tiny{}8} & {\tiny{}9} & {\tiny{}10} & {\tiny{}11} & {\tiny{}12}\tabularnewline
\hline 
{\tiny{}0} & {\tiny{}0.0000} & {\tiny{}0.3970} & {\tiny{}0.2336} & {\tiny{}0.1675} & {\tiny{}0.1227} & {\tiny{}0.0959} & {\tiny{}0.0902} & {\tiny{}0.0884} & {\tiny{}0.0997} & {\tiny{}0.0944} & {\tiny{}0.1020} & {\tiny{}0.0357} & {\tiny{}0.0370}\tabularnewline
{\tiny{}1} & {\tiny{}0.0726} & {\tiny{}0.3971} & {\tiny{}0.0323} & {\tiny{}0.0064} & {\tiny{}0.0033} & {\tiny{}0.0016} & {\tiny{}0.0019} & {\tiny{}0.0012} & {\tiny{}0.0012} & {\tiny{}0.0000} & {\tiny{}0.0000} & {\tiny{}0.0179} & {\tiny{}0.0000}\tabularnewline
{\tiny{}2} & {\tiny{}0.2776} & {\tiny{}0.1538} & {\tiny{}0.5566} & {\tiny{}0.0705} & {\tiny{}0.0124} & {\tiny{}0.0044} & {\tiny{}0.0017} & {\tiny{}0.0010} & {\tiny{}0.0000} & {\tiny{}0.0000} & {\tiny{}0.0000} & {\tiny{}0.0000} & {\tiny{}0.0000}\tabularnewline
{\tiny{}3} & {\tiny{}0.3012} & {\tiny{}0.0330} & {\tiny{}0.1407} & {\tiny{}0.6128} & {\tiny{}0.0801} & {\tiny{}0.0112} & {\tiny{}0.0054} & {\tiny{}0.0021} & {\tiny{}0.0012} & {\tiny{}0.0020} & {\tiny{}0.0000} & {\tiny{}0.0000} & {\tiny{}0.0000}\tabularnewline
{\tiny{}4} & {\tiny{}0.2344} & {\tiny{}0.0151} & {\tiny{}0.0305} & {\tiny{}0.1312} & {\tiny{}0.6958} & {\tiny{}0.1251} & {\tiny{}0.0193} & {\tiny{}0.0082} & {\tiny{}0.0031} & {\tiny{}0.0000} & {\tiny{}0.0000} & {\tiny{}0.0000} & {\tiny{}0.0000}\tabularnewline
{\tiny{}5} & {\tiny{}0.0713} & {\tiny{}0.0024} & {\tiny{}0.0049} & {\tiny{}0.0094} & {\tiny{}0.0770} & {\tiny{}0.6744} & {\tiny{}0.1155} & {\tiny{}0.0107} & {\tiny{}0.0055} & {\tiny{}0.0000} & {\tiny{}0.0102} & {\tiny{}0.0000} & {\tiny{}0.0000}\tabularnewline
\hline 
\end{tabular}{\tiny\par}
\par\end{center}

\noindent \begin{center}
Table 12 The first-order transfer matrix from 2005 to 2006
\par\end{center}

\noindent \begin{center}
{\tiny{}}%
\begin{tabular}{llllllllllllll}
\hline 
{\tiny{}Size} & {\tiny{}0} & {\tiny{}1} & {\tiny{}2} & {\tiny{}3} & {\tiny{}4} & {\tiny{}5} & {\tiny{}6} & {\tiny{}7} & {\tiny{}8} & {\tiny{}9} & {\tiny{}10} & {\tiny{}11} & {\tiny{}12}\tabularnewline
\hline 
{\tiny{}0} & {\tiny{}0.0000} & {\tiny{}0.2534} & {\tiny{}0.1374} & {\tiny{}0.1052} & {\tiny{}0.0811} & {\tiny{}0.0648} & {\tiny{}0.0505} & {\tiny{}0.0533} & {\tiny{}0.0584} & {\tiny{}0.0573} & {\tiny{}0.0183} & {\tiny{}0.0500} & {\tiny{}0.0000}\tabularnewline
{\tiny{}1} & {\tiny{}0.0765} & {\tiny{}0.5485} & {\tiny{}0.0343} & {\tiny{}0.0070} & {\tiny{}0.0033} & {\tiny{}0.0015} & {\tiny{}0.0011} & {\tiny{}0.0011} & {\tiny{}0.0000} & {\tiny{}0.0000} & {\tiny{}0.0000} & {\tiny{}0.0000} & {\tiny{}0.0000}\tabularnewline
{\tiny{}2} & {\tiny{}0.3049} & {\tiny{}0.1509} & {\tiny{}0.6777} & {\tiny{}0.0722} & {\tiny{}0.0131} & {\tiny{}0.0043} & {\tiny{}0.0021} & {\tiny{}0.0010} & {\tiny{}0.0000} & {\tiny{}0.0017} & {\tiny{}0.0000} & {\tiny{}0.0000} & {\tiny{}0.0000}\tabularnewline
{\tiny{}3} & {\tiny{}0.3046} & {\tiny{}0.0295} & {\tiny{}0.1247} & {\tiny{}0.7001} & {\tiny{}0.0792} & {\tiny{}0.0110} & {\tiny{}0.0054} & {\tiny{}0.0023} & {\tiny{}0.0000} & {\tiny{}0.0000} & {\tiny{}0.0000} & {\tiny{}0.0000} & {\tiny{}0.0000}\tabularnewline
{\tiny{}4} & {\tiny{}0.2226} & {\tiny{}0.0143} & {\tiny{}0.0227} & {\tiny{}0.1076} & {\tiny{}0.7593} & {\tiny{}0.1175} & {\tiny{}0.0205} & {\tiny{}0.0073} & {\tiny{}0.0017} & {\tiny{}0.0017} & {\tiny{}0.0046} & {\tiny{}0.0000} & {\tiny{}0.0000}\tabularnewline
{\tiny{}5} & {\tiny{}0.0610} & {\tiny{}0.0026} & {\tiny{}0.0027} & {\tiny{}0.0066} & {\tiny{}0.0592} & {\tiny{}0.7370} & {\tiny{}0.1108} & {\tiny{}0.0146} & {\tiny{}0.0040} & {\tiny{}0.0035} & {\tiny{}0.0000} & {\tiny{}0.0000} & {\tiny{}0.0000}\tabularnewline
\hline 
\end{tabular}{\tiny\par}
\par\end{center}

\noindent \begin{center}
Table 13 The first-order transfer matrix from 2006 to 2007
\par\end{center}

\noindent \begin{center}
{\tiny{}}%
\begin{tabular}{llllllllllllll}
\hline 
{\tiny{}Size} & {\tiny{}0} & {\tiny{}1} & {\tiny{}2} & {\tiny{}3} & {\tiny{}4} & {\tiny{}5} & {\tiny{}6} & {\tiny{}7} & {\tiny{}8} & {\tiny{}9} & {\tiny{}10} & {\tiny{}11} & {\tiny{}12}\tabularnewline
\hline 
{\tiny{}0} & {\tiny{}0.0000} & {\tiny{}0.2567} & {\tiny{}0.1320} & {\tiny{}0.0962} & {\tiny{}0.0749} & {\tiny{}0.0557} & {\tiny{}0.0486} & {\tiny{}0.0445} & {\tiny{}0.0429} & {\tiny{}0.0475} & {\tiny{}0.0364} & {\tiny{}0.0175} & {\tiny{}0.0000}\tabularnewline
{\tiny{}1} & {\tiny{}0.0784} & {\tiny{}0.5182} & {\tiny{}0.0294} & {\tiny{}0.0059} & {\tiny{}0.0029} & {\tiny{}0.0013} & {\tiny{}0.0011} & {\tiny{}0.0005} & {\tiny{}0.0000} & {\tiny{}0.0017} & {\tiny{}0.0000} & {\tiny{}0.0000} & {\tiny{}0.0000}\tabularnewline
{\tiny{}2} & {\tiny{}0.3235} & {\tiny{}0.1669} & {\tiny{}0.6846} & {\tiny{}0.0694} & {\tiny{}0.0124} & {\tiny{}0.0036} & {\tiny{}0.0015} & {\tiny{}0.0003} & {\tiny{}0.0000} & {\tiny{}0.0000} & {\tiny{}0.0000} & {\tiny{}0.0000} & {\tiny{}0.0000}\tabularnewline
{\tiny{}3} & {\tiny{}0.2972} & {\tiny{}0.0377} & {\tiny{}0.1247} & {\tiny{}0.7113} & {\tiny{}0.0774} & {\tiny{}0.0113} & {\tiny{}0.0061} & {\tiny{}0.0011} & {\tiny{}0.0000} & {\tiny{}0.0000} & {\tiny{}0.0000} & {\tiny{}0.0000} & {\tiny{}0.0000}\tabularnewline
{\tiny{}4} & {\tiny{}0.2142} & {\tiny{}0.0175} & {\tiny{}0.0249} & {\tiny{}0.1074} & {\tiny{}0.7642} & {\tiny{}0.1171} & {\tiny{}0.0198} & {\tiny{}0.0085} & {\tiny{}0.0016} & {\tiny{}0.0017} & {\tiny{}0.0000} & {\tiny{}0.0000} & {\tiny{}0.0000}\tabularnewline
{\tiny{}5} & {\tiny{}0.0578} & {\tiny{}0.0026} & {\tiny{}0.0038} & {\tiny{}0.0084} & {\tiny{}0.0630} & {\tiny{}0.7416} & {\tiny{}0.1119} & {\tiny{}0.0143} & {\tiny{}0.0071} & {\tiny{}0.0034} & {\tiny{}0.0000} & {\tiny{}0.0000} & {\tiny{}0.0000}\tabularnewline
\hline 
\end{tabular}{\tiny\par}
\par\end{center}

\pagebreak{}
\noindent \begin{center}
Table 14 The first-order transfer matrix from 2007 to 2008
\par\end{center}

\noindent \begin{center}
{\tiny{}}%
\begin{tabular}{llllllllllllll}
\hline 
{\tiny{}Size} & {\tiny{}0} & {\tiny{}1} & {\tiny{}2} & {\tiny{}3} & {\tiny{}4} & {\tiny{}5} & {\tiny{}6} & {\tiny{}7} & {\tiny{}8} & {\tiny{}9} & {\tiny{}10} & {\tiny{}11} & {\tiny{}12}\tabularnewline
\hline 
{\tiny{}0} & {\tiny{}0.0000} & {\tiny{}0.3513} & {\tiny{}0.2895} & {\tiny{}0.2654} & {\tiny{}0.2463} & {\tiny{}0.2314} & {\tiny{}0.2254} & {\tiny{}0.2339} & {\tiny{}0.2406} & {\tiny{}0.2324} & {\tiny{}0.2444} & {\tiny{}0.1525} & {\tiny{}0.2121}\tabularnewline
{\tiny{}1} & {\tiny{}0.0830} & {\tiny{}0.3699} & {\tiny{}0.0379} & {\tiny{}0.0058} & {\tiny{}0.0030} & {\tiny{}0.0011} & {\tiny{}0.0010} & {\tiny{}0.0004} & {\tiny{}0.0011} & {\tiny{}0.0000} & {\tiny{}0.0000} & {\tiny{}0.0000} & {\tiny{}0.0000}\tabularnewline
{\tiny{}2} & {\tiny{}0.2798} & {\tiny{}0.1998} & {\tiny{}0.4840} & {\tiny{}0.0956} & {\tiny{}0.0181} & {\tiny{}0.0053} & {\tiny{}0.0024} & {\tiny{}0.0012} & {\tiny{}0.0005} & {\tiny{}0.0016} & {\tiny{}0.0000} & {\tiny{}0.0000} & {\tiny{}0.0000}\tabularnewline
{\tiny{}3} & {\tiny{}0.2644} & {\tiny{}0.0508} & {\tiny{}0.1405} & {\tiny{}0.4989} & {\tiny{}0.1061} & {\tiny{}0.0150} & {\tiny{}0.0051} & {\tiny{}0.0024} & {\tiny{}0.0005} & {\tiny{}0.0000} & {\tiny{}0.0000} & {\tiny{}0.0000} & {\tiny{}0.0000}\tabularnewline
{\tiny{}4} & {\tiny{}0.2287} & {\tiny{}0.0256} & {\tiny{}0.0420} & {\tiny{}0.1205} & {\tiny{}0.5503} & {\tiny{}0.1566} & {\tiny{}0.0254} & {\tiny{}0.0095} & {\tiny{}0.0033} & {\tiny{}0.0016} & {\tiny{}0.0000} & {\tiny{}0.0000} & {\tiny{}0.0000}\tabularnewline
{\tiny{}5} & {\tiny{}0.0939} & {\tiny{}0.0021} & {\tiny{}0.0056} & {\tiny{}0.0122} & {\tiny{}0.0688} & {\tiny{}0.5241} & {\tiny{}0.1425} & {\tiny{}0.0185} & {\tiny{}0.0087} & {\tiny{}0.0000} & {\tiny{}0.0038} & {\tiny{}0.0000} & {\tiny{}0.0000}\tabularnewline
\hline 
\end{tabular}{\tiny\par}
\par\end{center}

\noindent \begin{center}
Table 15 The first-order transfer matrix from 2008 to 2009
\par\end{center}

\noindent \begin{center}
{\tiny{}}%
\begin{tabular}{llllllllllllll}
\hline 
{\tiny{}Size} & {\tiny{}0} & {\tiny{}1} & {\tiny{}2} & {\tiny{}3} & {\tiny{}4} & {\tiny{}5} & {\tiny{}6} & {\tiny{}7} & {\tiny{}8} & {\tiny{}9} & {\tiny{}10} & {\tiny{}11} & {\tiny{}12}\tabularnewline
\hline 
{\tiny{}0} & {\tiny{}0.0000} & {\tiny{}0.3340} & {\tiny{}0.2711} & {\tiny{}0.2476} & {\tiny{}0.2365} & {\tiny{}0.2324} & {\tiny{}0.2216} & {\tiny{}0.2147} & {\tiny{}0.2391} & {\tiny{}0.2188} & {\tiny{}0.1690} & {\tiny{}0.1250} & {\tiny{}0.1923}\tabularnewline
{\tiny{}1} & {\tiny{}0.0843} & {\tiny{}0.4809} & {\tiny{}0.0397} & {\tiny{}0.0067} & {\tiny{}0.0034} & {\tiny{}0.0014} & {\tiny{}0.0007} & {\tiny{}0.0013} & {\tiny{}0.0000} & {\tiny{}0.0020} & {\tiny{}0.0000} & {\tiny{}0.0000} & {\tiny{}0.0000}\tabularnewline
{\tiny{}2} & {\tiny{}0.3238} & {\tiny{}0.1486} & {\tiny{}0.5699} & {\tiny{}0.0984} & {\tiny{}0.0196} & {\tiny{}0.0062} & {\tiny{}0.0033} & {\tiny{}0.0018} & {\tiny{}0.0020} & {\tiny{}0.0000} & {\tiny{}0.0000} & {\tiny{}0.0000} & {\tiny{}0.0000}\tabularnewline
{\tiny{}3} & {\tiny{}0.2821} & {\tiny{}0.0230} & {\tiny{}0.0960} & {\tiny{}0.5582} & {\tiny{}0.1064} & {\tiny{}0.0173} & {\tiny{}0.0057} & {\tiny{}0.0050} & {\tiny{}0.0020} & {\tiny{}0.0000} & {\tiny{}0.0000} & {\tiny{}0.0000} & {\tiny{}0.0000}\tabularnewline
{\tiny{}4} & {\tiny{}0.1985} & {\tiny{}0.0106} & {\tiny{}0.0184} & {\tiny{}0.0787} & {\tiny{}0.5769} & {\tiny{}0.1514} & {\tiny{}0.0245} & {\tiny{}0.0116} & {\tiny{}0.0048} & {\tiny{}0.0000} & {\tiny{}0.0000} & {\tiny{}0.0000} & {\tiny{}0.0000}\tabularnewline
{\tiny{}5} & {\tiny{}0.0692} & {\tiny{}0.0021} & {\tiny{}0.0040} & {\tiny{}0.0089} & {\tiny{}0.0526} & {\tiny{}0.5406} & {\tiny{}0.1488} & {\tiny{}0.0164} & {\tiny{}0.0041} & {\tiny{}0.0000} & {\tiny{}0.0047} & {\tiny{}0.0000} & {\tiny{}0.0000}\tabularnewline
\hline 
\end{tabular}{\tiny\par}
\par\end{center}

\noindent \begin{center}
Table 16 The first-order transfer matrix from 2009 to 2010
\par\end{center}

\noindent \begin{center}
{\tiny{}}%
\begin{tabular}{llllllllllllll}
\hline 
{\tiny{}Size} & {\tiny{}0} & {\tiny{}1} & {\tiny{}2} & {\tiny{}3} & {\tiny{}4} & {\tiny{}5} & {\tiny{}6} & {\tiny{}7} & {\tiny{}8} & {\tiny{}9} & {\tiny{}10} & {\tiny{}11} & {\tiny{}12}\tabularnewline
\hline 
{\tiny{}0} & {\tiny{}0.0000} & {\tiny{}0.3499} & {\tiny{}0.3060} & {\tiny{}0.2341} & {\tiny{}0.1740} & {\tiny{}0.1368} & {\tiny{}0.1041} & {\tiny{}0.0926} & {\tiny{}0.0877} & {\tiny{}0.0863} & {\tiny{}0.0635} & {\tiny{}0.0392} & {\tiny{}0.1379}\tabularnewline
{\tiny{}1} & {\tiny{}0.0410} & {\tiny{}0.3177} & {\tiny{}0.0450} & {\tiny{}0.0089} & {\tiny{}0.0043} & {\tiny{}0.0015} & {\tiny{}0.0007} & {\tiny{}0.0004} & {\tiny{}0.0008} & {\tiny{}0.0000} & {\tiny{}0.0000} & {\tiny{}0.0000} & {\tiny{}0.0000}\tabularnewline
{\tiny{}2} & {\tiny{}0.1737} & {\tiny{}0.2284} & {\tiny{}0.4389} & {\tiny{}0.1204} & {\tiny{}0.0360} & {\tiny{}0.0171} & {\tiny{}0.0039} & {\tiny{}0.0023} & {\tiny{}0.0017} & {\tiny{}0.0000} & {\tiny{}0.0000} & {\tiny{}0.0000} & {\tiny{}0.0000}\tabularnewline
{\tiny{}3} & {\tiny{}0.2119} & {\tiny{}0.0612} & {\tiny{}0.1641} & {\tiny{}0.4674} & {\tiny{}0.1351} & {\tiny{}0.0435} & {\tiny{}0.0130} & {\tiny{}0.0069} & {\tiny{}0.0000} & {\tiny{}0.0000} & {\tiny{}0.0000} & {\tiny{}0.0000} & {\tiny{}0.0000}\tabularnewline
{\tiny{}4} & {\tiny{}0.2132} & {\tiny{}0.0324} & {\tiny{}0.0393} & {\tiny{}0.1538} & {\tiny{}0.5467} & {\tiny{}0.2106} & {\tiny{}0.0553} & {\tiny{}0.0200} & {\tiny{}0.0066} & {\tiny{}0.0126} & {\tiny{}0.0053} & {\tiny{}0.0000} & {\tiny{}0.0000}\tabularnewline
{\tiny{}5} & {\tiny{}0.1890} & {\tiny{}0.0081} & {\tiny{}0.0051} & {\tiny{}0.0124} & {\tiny{}0.0934} & {\tiny{}0.4929} & {\tiny{}0.1872} & {\tiny{}0.0438} & {\tiny{}0.0215} & {\tiny{}0.0147} & {\tiny{}0.0053} & {\tiny{}0.0000} & {\tiny{}0.0000}\tabularnewline
\hline 
\end{tabular}{\tiny\par}
\par\end{center}

\pagebreak{}
\noindent \begin{center}
Table 17 The first-order transfer matrix from 2010 to 2011
\par\end{center}

\noindent \begin{center}
{\tiny{}}%
\begin{tabular}{llllllllllllll}
\hline 
{\tiny{}Size} & {\tiny{}0} & {\tiny{}1} & {\tiny{}2} & {\tiny{}3} & {\tiny{}4} & {\tiny{}5} & {\tiny{}6} & {\tiny{}7} & {\tiny{}8} & {\tiny{}9} & {\tiny{}10} & {\tiny{}11} & {\tiny{}12}\tabularnewline
\hline 
{\tiny{}0} & {\tiny{}0.0000} & {\tiny{}0.7743} & {\tiny{}0.6891} & {\tiny{}0.6076} & {\tiny{}0.4969} & {\tiny{}0.5833} & {\tiny{}0.5908} & {\tiny{}0.5649} & {\tiny{}0.5683} & {\tiny{}0.5741} & {\tiny{}0.5281} & {\tiny{}0.4752} & {\tiny{}0.6143}\tabularnewline
{\tiny{}1} & {\tiny{}0.0167} & {\tiny{}0.0057} & {\tiny{}0.0060} & {\tiny{}0.0065} & {\tiny{}0.0075} & {\tiny{}0.0062} & {\tiny{}0.0063} & {\tiny{}0.0059} & {\tiny{}0.0055} & {\tiny{}0.0066} & {\tiny{}0.0065} & {\tiny{}0.0099} & {\tiny{}0.0000}\tabularnewline
{\tiny{}2} & {\tiny{}0.0386} & {\tiny{}0.0191} & {\tiny{}0.0157} & {\tiny{}0.0119} & {\tiny{}0.0123} & {\tiny{}0.0097} & {\tiny{}0.0087} & {\tiny{}0.0094} & {\tiny{}0.0091} & {\tiny{}0.0094} & {\tiny{}0.0022} & {\tiny{}0.0099} & {\tiny{}0.0000}\tabularnewline
{\tiny{}3} & {\tiny{}0.1150} & {\tiny{}0.0438} & {\tiny{}0.0546} & {\tiny{}0.0419} & {\tiny{}0.0270} & {\tiny{}0.0169} & {\tiny{}0.0163} & {\tiny{}0.0160} & {\tiny{}0.0121} & {\tiny{}0.0113} & {\tiny{}0.0411} & {\tiny{}0.0297} & {\tiny{}0.0286}\tabularnewline
{\tiny{}4} & {\tiny{}0.4993} & {\tiny{}0.1354} & {\tiny{}0.1891} & {\tiny{}0.2277} & {\tiny{}0.1966} & {\tiny{}0.0696} & {\tiny{}0.0394} & {\tiny{}0.0372} & {\tiny{}0.0414} & {\tiny{}0.0415} & {\tiny{}0.0368} & {\tiny{}0.0594} & {\tiny{}0.0429}\tabularnewline
{\tiny{}5} & {\tiny{}0.2544} & {\tiny{}0.0203} & {\tiny{}0.0422} & {\tiny{}0.0961} & {\tiny{}0.2263} & {\tiny{}0.1950} & {\tiny{}0.0636} & {\tiny{}0.0145} & {\tiny{}0.0033} & {\tiny{}0.0009} & {\tiny{}0.0022} & {\tiny{}0.0000} & {\tiny{}0.0000}\tabularnewline
\hline 
\end{tabular}{\tiny\par}
\par\end{center}

\noindent \begin{center}
Table 18 The first-order transfer matrix from 2011 to 2012
\par\end{center}

\noindent \begin{center}
{\tiny{}}%
\begin{tabular}{llllllllllllll}
\hline 
{\tiny{}Size} & {\tiny{}0} & {\tiny{}1} & {\tiny{}2} & {\tiny{}3} & {\tiny{}4} & {\tiny{}5} & {\tiny{}6} & {\tiny{}7} & {\tiny{}8} & {\tiny{}9} & {\tiny{}10} & {\tiny{}11} & {\tiny{}12}\tabularnewline
\hline 
{\tiny{}0} & {\tiny{}0.0000} & {\tiny{}0.0014} & {\tiny{}0.0016} & {\tiny{}0.0020} & {\tiny{}0.0017} & {\tiny{}0.0013} & {\tiny{}0.0007} & {\tiny{}0.0006} & {\tiny{}0.0000} & {\tiny{}0.0000} & {\tiny{}0.0000} & {\tiny{}0.0000} & {\tiny{}0.0000}\tabularnewline
{\tiny{}1} & {\tiny{}0.0439} & {\tiny{}0.9986} & {\tiny{}0.0000} & {\tiny{}0.0000} & {\tiny{}0.0000} & {\tiny{}0.0000} & {\tiny{}0.0000} & {\tiny{}0.0000} & {\tiny{}0.0000} & {\tiny{}0.0000} & {\tiny{}0.0000} & {\tiny{}0.0000} & {\tiny{}0.0000}\tabularnewline
{\tiny{}2} & {\tiny{}0.0995} & {\tiny{}0.0000} & {\tiny{}0.9984} & {\tiny{}0.0000} & {\tiny{}0.0000} & {\tiny{}0.0000} & {\tiny{}0.0000} & {\tiny{}0.0000} & {\tiny{}0.0000} & {\tiny{}0.0000} & {\tiny{}0.0000} & {\tiny{}0.0000} & {\tiny{}0.0000}\tabularnewline
{\tiny{}3} & {\tiny{}0.1990} & {\tiny{}0.0000} & {\tiny{}0.0000} & {\tiny{}0.9896} & {\tiny{}0.0000} & {\tiny{}0.0000} & {\tiny{}0.0000} & {\tiny{}0.0000} & {\tiny{}0.0000} & {\tiny{}0.0000} & {\tiny{}0.0000} & {\tiny{}0.0000} & {\tiny{}0.0000}\tabularnewline
{\tiny{}4} & {\tiny{}0.5573} & {\tiny{}0.0000} & {\tiny{}0.0000} & {\tiny{}0.0084} & {\tiny{}0.9938} & {\tiny{}0.0000} & {\tiny{}0.0000} & {\tiny{}0.0000} & {\tiny{}0.0000} & {\tiny{}0.0000} & {\tiny{}0.0000} & {\tiny{}0.0000} & {\tiny{}0.0000}\tabularnewline
{\tiny{}5} & {\tiny{}0.0816} & {\tiny{}0.0000} & {\tiny{}0.0000} & {\tiny{}0.0000} & {\tiny{}0.0045} & {\tiny{}0.9958} & {\tiny{}0.0000} & {\tiny{}0.0000} & {\tiny{}0.0000} & {\tiny{}0.0000} & {\tiny{}0.0000} & {\tiny{}0.0000} & {\tiny{}0.0000}\tabularnewline
\hline 
\end{tabular}{\tiny\par}
\par\end{center}

\noindent \begin{center}
Table 19 The first-order transfer matrix from 2012 to 2013
\par\end{center}

\noindent \begin{center}
{\tiny{}}%
\begin{tabular}{llllllllllllll}
\hline 
{\tiny{}Size} & {\tiny{}0} & {\tiny{}1} & {\tiny{}2} & {\tiny{}3} & {\tiny{}4} & {\tiny{}5} & {\tiny{}6} & {\tiny{}7} & {\tiny{}8} & {\tiny{}9} & {\tiny{}10} & {\tiny{}11} & {\tiny{}12}\tabularnewline
\hline 
{\tiny{}0} & {\tiny{}0.0000} & {\tiny{}0.6480} & {\tiny{}0.5886} & {\tiny{}0.5510} & {\tiny{}0.5192} & {\tiny{}0.4873} & {\tiny{}0.4447} & {\tiny{}0.3886} & {\tiny{}0.3427} & {\tiny{}0.3101} & {\tiny{}0.2605} & {\tiny{}0.2055} & {\tiny{}0.1250}\tabularnewline
{\tiny{}1} & {\tiny{}0.0255} & {\tiny{}0.0150} & {\tiny{}0.0121} & {\tiny{}0.0079} & {\tiny{}0.0038} & {\tiny{}0.0000} & {\tiny{}0.0000} & {\tiny{}0.0000} & {\tiny{}0.0000} & {\tiny{}0.0000} & {\tiny{}0.0000} & {\tiny{}0.0000} & {\tiny{}0.0000}\tabularnewline
{\tiny{}2} & {\tiny{}0.0517} & {\tiny{}0.0257} & {\tiny{}0.0286} & {\tiny{}0.0192} & {\tiny{}0.0110} & {\tiny{}0.0000} & {\tiny{}0.0000} & {\tiny{}0.0000} & {\tiny{}0.0000} & {\tiny{}0.0000} & {\tiny{}0.0000} & {\tiny{}0.0000} & {\tiny{}0.0000}\tabularnewline
{\tiny{}3} & {\tiny{}0.0997} & {\tiny{}0.0402} & {\tiny{}0.0492} & {\tiny{}0.0511} & {\tiny{}0.0338} & {\tiny{}0.0023} & {\tiny{}0.0000} & {\tiny{}0.0000} & {\tiny{}0.0000} & {\tiny{}0.0000} & {\tiny{}0.0000} & {\tiny{}0.0000} & {\tiny{}0.0000}\tabularnewline
{\tiny{}4} & {\tiny{}0.3372} & {\tiny{}0.1272} & {\tiny{}0.1491} & {\tiny{}0.1637} & {\tiny{}0.1636} & {\tiny{}0.0722} & {\tiny{}0.0000} & {\tiny{}0.0000} & {\tiny{}0.0000} & {\tiny{}0.0000} & {\tiny{}0.0000} & {\tiny{}0.0000} & {\tiny{}0.0000}\tabularnewline
{\tiny{}5} & {\tiny{}0.4385} & {\tiny{}0.1438} & {\tiny{}0.1723} & {\tiny{}0.2071} & {\tiny{}0.2535} & {\tiny{}0.2837} & {\tiny{}0.0443} & {\tiny{}0.0000} & {\tiny{}0.0000} & {\tiny{}0.0000} & {\tiny{}0.0000} & {\tiny{}0.0000} & {\tiny{}0.0000}\tabularnewline
\hline 
\end{tabular}{\tiny\par}
\par\end{center}

\pagebreak{}

\subsection*{Obtained Transition Trend and Entropy Results}

The calculated transition trend and transition entropy of each year/category
are presented as follows:
\noindent \begin{center}
Table 20 The transition trend from 1999 to 2013
\par\end{center}

\noindent \begin{center}
\begin{tabular}{lllllllll}
\hline 
{\tiny{}Year{*}} & {\tiny{}1999} & {\tiny{}2000} & {\tiny{}2001} & {\tiny{}2002} & {\tiny{}2003} & {\tiny{}2004} & {\tiny{}2005} & {\tiny{}2006}\tabularnewline
\hline 
{\tiny{}L(d)} & {\tiny{}0.2825} & {\tiny{}0.2268} & {\tiny{}0.3446} & {\tiny{}0.2661} & {\tiny{}0.2889} & {\tiny{}0.4667} & {\tiny{}0.2396} & {\tiny{}0.2534}\tabularnewline
{\tiny{}R(d)} & {\tiny{}0.2500} & {\tiny{}0.2744} & {\tiny{}0.2791} & {\tiny{}0.2077} & {\tiny{}0.2092} & {\tiny{}0.2173} & {\tiny{}0.2144} & {\tiny{}0.1526}\tabularnewline
{\tiny{}Q(d)} & {\tiny{}1.1300} & {\tiny{}0.8265} & {\tiny{}1.2347} & {\tiny{}1.2814} & {\tiny{}1.3811} & {\tiny{}2.1482} & {\tiny{}1.1179} & {\tiny{}1.6603}\tabularnewline
\cline{1-8} \cline{2-8} \cline{3-8} \cline{4-8} \cline{5-8} \cline{6-8} \cline{7-8} \cline{8-8} 
{\tiny{}Year{*}} & {\tiny{}2007} & {\tiny{}2008} & {\tiny{}2009} & {\tiny{}2010} & {\tiny{}2011} & {\tiny{}2012} & {\tiny{}2013} & \tabularnewline
\cline{1-8} \cline{2-8} \cline{3-8} \cline{4-8} \cline{5-8} \cline{6-8} \cline{7-8} \cline{8-8} 
{\tiny{}L(d)} & {\tiny{}0.2609} & {\tiny{}0.1289} & {\tiny{}0.2185} & {\tiny{}0.5030} & {\tiny{}0.4439} & {\tiny{}0.1344} & {\tiny{}0.3185} & \tabularnewline
{\tiny{}R(d)} & {\tiny{}0.1448} & {\tiny{}0.3643} & {\tiny{}0.3045} & {\tiny{}0.2122} & {\tiny{}0.4695} & {\tiny{}0.0313} & {\tiny{}0.4952} & \tabularnewline
{\tiny{}Q(d)} & {\tiny{}1.8020} & {\tiny{}0.3538} & {\tiny{}0.7174} & {\tiny{}2.3700} & {\tiny{}0.9455} & {\tiny{}4.2973} & {\tiny{}0.6432} & \tabularnewline
\cline{1-8} \cline{2-8} \cline{3-8} \cline{4-8} \cline{5-8} \cline{6-8} \cline{7-8} \cline{8-8} 
\multicolumn{9}{r}{{\tiny{}{*}For simplicity, year in this table means end year. e.g.,
year 1999 represents 1998 to 1999.}}\tabularnewline
\end{tabular}
\par\end{center}

\noindent \begin{center}
Table 21 The transition entropy from 1999 to 2013
\par\end{center}

\noindent \begin{center}
{\scriptsize{}}%
\begin{tabular}{llllllll|l}
\hline 
{\tiny{}Category} & {\tiny{}1999} & {\tiny{}2000} & {\tiny{}2001} & {\tiny{}2002} & {\tiny{}2003} & {\tiny{}2004} & \multicolumn{1}{l}{{\tiny{}2005}} & {\tiny{}2006}\tabularnewline
\hline 
{\tiny{}0} & {\tiny{}1.7736} & {\tiny{}1.7277} & {\tiny{}1.7185} & {\tiny{}1.6733} & {\tiny{}1.6278} & {\tiny{}1.6290} & \multicolumn{1}{l}{{\tiny{}1.6127}} & {\tiny{}1.5613}\tabularnewline
{\tiny{}1} & {\tiny{}1.3928} & {\tiny{}1.1951} & {\tiny{}1.5558} & {\tiny{}1.1599} & {\tiny{}1.1458} & {\tiny{}1.1342} & \multicolumn{1}{l}{{\tiny{}1.2228}} & {\tiny{}1.1497}\tabularnewline
{\tiny{}2} & {\tiny{}1.3500} & {\tiny{}1.2285} & {\tiny{}1.3404} & {\tiny{}1.1952} & {\tiny{}1.1900} & {\tiny{}1.2450} & \multicolumn{1}{l}{{\tiny{}1.1945}} & {\tiny{}1.0183}\tabularnewline
{\tiny{}3} & {\tiny{}1.2867} & {\tiny{}1.2593} & {\tiny{}1.3220} & {\tiny{}1.1629} & {\tiny{}1.1601} & {\tiny{}1.2444} & \multicolumn{1}{l}{{\tiny{}1.1441}} & {\tiny{}0.9938}\tabularnewline
{\tiny{}4} & {\tiny{}1.1261} & {\tiny{}1.1642} & {\tiny{}1.2224} & {\tiny{}1.0384} & {\tiny{}1.0507} & {\tiny{}1.1789} & \multicolumn{1}{l}{{\tiny{}1.0281}} & {\tiny{}0.8853}\tabularnewline
{\tiny{}5} & {\tiny{}1.1603} & {\tiny{}1.2119} & {\tiny{}1.2745} & {\tiny{}1.0764} & {\tiny{}1.0954} & {\tiny{}1.2498} & \multicolumn{1}{l}{{\tiny{}1.0754}} & {\tiny{}0.9278}\tabularnewline
{\tiny{}6} & {\tiny{}1.1683} & {\tiny{}1.1962} & {\tiny{}1.2808} & {\tiny{}1.0500} & {\tiny{}1.0657} & {\tiny{}1.2618} & \multicolumn{1}{l}{{\tiny{}1.0817}} & {\tiny{}0.9123}\tabularnewline
\hline 
\end{tabular}{\scriptsize\par}
\par\end{center}

\end{document}